\documentclass[a4paper, 10pt]{llncs}

\begin{document}

\title{An ExpTime Procedure for  Description Logic $\mathcal{ALCQI}$ (Draft)}
\author{Yu Ding}

\institute{Concordia University, Montreal QC H3G 1M8, Canada\\
\email{ding\_yu@cse.concordia.ca} \\ Document Created: Aug-10-2006 \\ Latest Modification: Jan-11-2007  }

\pagestyle{plain}

\maketitle

\begin{abstract}  A worst-case ExpTime tableau-based decision procedure is outlined for the satisfiability problem in $\mathcal{ALCQI}$ w.r.t. general axioms. \end{abstract}

\section{Motivation and Brief Introduction}

The \emph{concept satisfiability problem} in description logics (DLs) with  both $\mathcal{Q}$ and $\mathcal{I}$ has been  considered empirically the hardest of all for those DL problems in the ExpTime complexity class.  Though the C-rule (the Ramsey's Rule)\cite{Ding2007b} works for other logics like $\mathcal{ALCFI}$ or $\mathcal{ALCOI}$, it is not obviously applicable to DLs with the \emph{qualified number restrictions}. In this paper, we take a different and general approach for $\mathcal{ALCQI}$. The focus is an ExpTime tableau-based  procedure and therefore empirical issues are not concerned. We start with a  brief introduction to the DL $\mathcal{ALCQI}$, the \emph{general inclusion axioms}, and the \emph{concept satisfiability  problem}. For more  we refer to \cite{DLhbk03}.

\begin{definition} \emph{\textbf{(Concept Formulae)}} We use $A$ for \emph{atomic concept}, use $C$ and $D$ for arbitrary concepts, use $R$ for a role name.  For non-negative integer $n$, concept formulae in $\mathcal{ALCQI}$ are formed according the following grammar\footnote{W.l.o.g. $\exists R.C$ is expressed in $\exists^{\geq 1} R.C$, and $\forall R.\neg C$ is expressed in $\exists^{\leq 0} R.C$.}:

\noindent $C,D :=  \top | A | \neg C | C \sqcap D| C \sqcup D| \exists^{\leq n} R.C | \exists^{\geq n} R.C$  \end{definition}

\begin{definition} \emph{\textbf{(Semantics)}} An interpretation $\mathcal{I} = (\Delta^\mathcal{I}, .^\mathcal{I})$ consists of a set $\Delta^\mathcal{I}$ (the domain) and an interpretation function $.^\mathcal{I}$. The interpretation function maps each concept name $C$ to a subset $C^\mathcal{I}$ of $\Delta^\mathcal{I}$, each role name $R$ to a subset $R^\mathcal{I}$ of $\Delta^\mathcal{I} \times \Delta^\mathcal{I}$. Let the symbols $C, D$ be concept formulae, $R$ be a role name. The interpretation function can be inductively defined as follows:

\noindent $\top^\mathcal{I} := \Delta^\mathcal{I}$  \hspace{4cm} $(\neg C)^\mathcal{I} := \Delta^\mathcal{I} \setminus C^\mathcal{I}$

\noindent $(C \sqcap D)^\mathcal{I} := C^\mathcal{I} \cap D^\mathcal{I}$  \hspace{2cm} \ $(C \sqcup D)^\mathcal{I} := C^\mathcal{I} \cup D^\mathcal{I}$

\noindent $(\exists^{\leq n} R.C)^\mathcal{I} := \{x \in \Delta^\mathcal{I} \  | \  ||\{ y \in \Delta^\mathcal{I}: (x, y) \in R^\mathcal{I}$ and $y \in C^\mathcal{I}\}||\leq n\}$

\noindent $(\exists^{\geq n} R.C)^\mathcal{I} := \{x \in \Delta^\mathcal{I} \ | \ ||\{y \in \Delta^\mathcal{I}: (x, y) \in R^\mathcal{I}$ and $y \in C^\mathcal{I}\}||\geq n \}$

\noindent and additionally, it satisfies  $(x, y) \in R^\mathcal{I} \Leftrightarrow (y, x) \in (R^-)^\mathcal{I}$. \end{definition}

\begin{definition} \emph{\textbf{(Negation Norm Form)}} The negation normal form is defined by applying the following transformation in such a way that negation signs are pushed inward and appear only in front of concept names.

\noindent  $\neg \neg C \rightarrow C$                  \hspace{4cm}  $\neg(C \sqcap D) \rightarrow \neg C \sqcup \neg D$

\noindent  $\neg(C \sqcup D) \rightarrow \neg C \sqcap \neg D$ \hspace{2cm}  \  $\neg \exists^{\leq n} R.C \rightarrow \exists^{\geq n + 1} R.C$

\noindent  $\neg \exists^{\geq n} R.C \rightarrow \exists^{\leq n-1} R.C$  \end{definition}

\begin{definition} \emph{\textbf{(Generalized Concept Inclusions)}} If $C$ is a concept formula, then $\top \sqsubseteq C$ (generalized concept inclusion or GCI) is a terminological axioms. A finite set of terminological axioms $\mathcal{T}$ is called a Tbox. The interpretation function $.^\mathcal{I}$ is extended for GCI as $(\top \sqsubseteq C)^\mathcal{I} := \top^\mathcal{I} \subseteq C^\mathcal{I}$. Without lose of generality, the general inclusion axioms can be expressed in one bigger GCI in NNF.   \end{definition}

\section{Preliminaries and Notations}

In the paper\footnote{For
brevity, $\exists^{\triangleright\!\triangleleft n}R.C$ denotes
$\exists^{\leq n}R.C$ or $\exists^{\geq n}R.C$, and $\tilde{C}$ is the NNF of $\neg C$.}  we call $\exists^{\leq n}
R.C$ and $\exists^{\geq n} R.C$ \emph{modal constraints}, and call
$C \sqcap D$ and $C \sqcup D$ \emph{propositional constraints}.  We assume each role has a unique inverse role. For a role $R$,  for
example, we consider $R^-$ as the  only \emph{inverse
role}\footnote{It takes a linear cost to identify equivalent role
names that are implied by the declarations of inverse relationship
in a  namespace (of role names).}.  

The discussion is put in the context of labeled trees. Each node is labeled with a set of 
concept formulae, each edge is labeled with a role\footnote{The inverse relationship can be ignored due
to the cut formulae introduced below.}. What is important is to each (tableau-tree) node $x$ we also attach algebraic objects like 
\emph{systems of linear integer inequalities} (LIIs) $lii(x, R)$, and to each $R$-edge (to $x$'s successors) we attach one \emph{non-negative integer solution} $S(x, R)$ of $lii(x, R)$.

We basically require that readers are familiar with \emph{propositional
logic} and \emph{integer linear programming}\cite{linbook}\cite{schrijver86} (plus a bit knowledge
of integer \emph{matrix} and \emph{linear algebra}).  Several notions are
to be explained below.

\subsection{Cut Formulae}

\begin{definition} \emph{\textbf{(Cut Formulae)}} Give a concept $E$ and a GCI $G$ in $\mathcal{ALCQI}$ for satisfiability test. For each modal subformula of $G$ and $E$ of the form $\exists^{\triangleright\!\triangleleft n}R.C$, where $R$ is any role, there is one cut formula as: 

 ($\star$) $\exists^{\leq 0}R^-.\top \sqcup C \sqcup \tilde{C}$. 

\noindent The set of all cut formulae for $E$ and $G$ is denoted as $\mathcal{K}_a$. \end{definition}

The set $\mathcal{K}_a$ is trivially satisfiable\footnote{To be precise, any model for $E$
and $G$ can be extended to satisfy $\mathcal{K}_a$.} in any model
for $E$ and $G$. The most important to notice is that, due to  the
cut-formulae, the calculus can treat $R^-$ and $R$ as independent
role names as if they had no inverse relationship at all. When this is 
exploited in the tree-like tableaux structure, the construction can be performed 
top-down and each node will be visited only once.

We denote as $\mathcal{C}(x)$ the result from splitting the cut formulae at $x$'s $R^-$-predecessor node and simply call it the \emph{cut-set} for $x$. 
For a cut-formula  $\exists^{\leq 0}R^-.\top \sqcup D \sqcup \tilde{D}$ at $x$'s $R^-$-predecessor, either $D \in \mathcal{C}(x)$ or $\tilde{D} \in \mathcal{C}(x)$. 

\subsection{Propositional Branch and Its Fine Tune}

\begin{definition} \emph{\textbf{(Propositional Branches)}}  Give  $\mathcal{L}(x)$ the set of labels for the node/element $x$, the propositional branches (PBs) for $x$  is  $\mathcal{BS}(x)$  the set of all possible disjuncts from the disjunctive normal form\footnote{We do not need a canonical (propositional) form and therefore DNF suffices. We treat each propositional branch as a set of modal constraints or concept literals.} (DNF) of $\mathcal{L}(x)$ by treating modal constraints as propositions. Denote the finite set of PBs as $\mathcal{BS}(x) = \{\mathcal{B}_1(x), ..., \mathcal{B}_i(x) ...\}$. \end{definition}

The notion of \emph{propositional branches} (PBs) is quite intuitive if one considers the AND-OR structure of concept formulae and the results from exhaustively performing the $\sqcap$-rule and $\sqcup$-rule commonly seen in tableaux calculi such as for $\mathcal{ALC}$. Enumerating PBs for a set of labels means handling all outer $\sqcap$ and $\sqcup$ operators in this AND-OR structure (other than those located inside role fillers).

\begin{definition} \emph{\textbf{(Fine-Tuned Modal Constraints)}}
In the tableaux (labeled tree) \texttt{T}, let $x$ be the $R^-$-predecessor of $y$,  we have: \begin{itemize}
\item Give $x \in C^\mathcal{I}$,  then $y \in (\exists^{\leq n}R.C)^\mathcal{I}$ iff $\| \{ z \in \Delta^\mathcal{I}: (y, z) \in R^\mathcal{I}$ and $z \in C^\mathcal{I}$ and $z$ is $R$-successor of $ y \} \| \leq n - 1$;
\item Give $x \in C^\mathcal{I}$, then $y \in (\exists^{\geq n}R.C)^\mathcal{I}$ iff $\| \{ z \in \Delta^\mathcal{I}: (y, z) \in R^\mathcal{I}$ and $z \in C^\mathcal{I}$ and $z$ is $R$-successor of $ y \} \| \geq n - 1$;\end{itemize} 
These adjustments of cardinalities over successors depending on the cut-set chosen at the predecessor are called fine-tuning of modal constraints. We denote the propositional branch $\mathcal{B}(x)$ after fine-tuning as $\mathcal{B}'(x)$.  \end{definition} 

\subsection{Linear Diophantine Inequalities}

The procedure will be presented as in the \emph{algebraic approach}. We reuse the \emph{atomic decomposition} technique. What is typical of the algebraic approach\footnote{\label{x}Regardless of the differences, the \textsl{atomic decomposition} and the special \textsl{linear integer inequalities} have intricate connections to the \textsl{choose-rule} and Tobies's \textsl{counter}.} is the building of systems of LIIs from decompositions of role fillers on each role.  For more we refer to Ohlbach's\cite{Ohlbach99}, Haarslev and M$\ddot{o}$ller's\cite{Haarslev01comb} work.

\begin{definition} \emph{\textbf{(Linear Integer Inequalities)}} Linear (subset sum)
 integer inequalities (LII) is a system of special linear Diophantine inequalities (LDI) such that,
 for the finite set of variables $V = \{v_1, v_2, \ ..., \ v_j, \ ..., v_{2^\lambda - 1}\}$ from the
 non-negative integer domain, the $k$-th LDI is of the form $(\sum^{2^\lambda - 1}_{j=1} v_j \cdot w_{k,j}) \leq n_k$
 or of the form $(\sum^{2^\lambda - 1}_{j=1}  v_j \cdot w_{k,j}) \geq n_k$, where each constant $w_{k, j} \in \{0, 1\}$, each
 unknown variable $v_j \in V$ is in the non-negative integer domain, and each $n_k$ is some
 non-negative integer constant, $\lambda$ is a non-negative integer constant.
 The number of unknown variables is $2^\lambda - 1$ where $\lambda$ is the number of LDIs and
 is also the number of modal constraints before atom-decomposition. \end{definition}

\section{The Decision Procedure for  $\mathcal{ALCQI}$}

  For (tableau-tree) node $x$, we use $\mathcal{L}(x)$ for its initial label, and
$\mathcal{B}(x)$ for its current \emph{propositional branch}, and
$\mathcal{B}'(x)$ for the corresponding \emph{fine-tuned} one. The converted problem is $E$ and $G \cup \mathcal{K}_a$ (in which the very special \emph{cut-formulae} are contained). Below is a set of  expansion rules for the converted problem.

\vspace{0.2cm}

        \begin{tabular}{cll} \hline

          $PB$-rule: &  if & 1. $x$ is not blocked, and $x$ is an $R$-successor, and \\
   &  & 2.  $\left\langle \mathcal{C}(x), R, \mathcal{L}(x) \right\rangle \notin $ \texttt{Nogood}, and \\
   &  & 3.  there is a $\mathcal{B}(x) \in \mathcal{BS}(x)$ such that\\
   &  & \hspace{0.4cm} (a) $\{\left\langle \emptyset, \epsilon, \mathcal{B}(x) \right\rangle, \left\langle \emptyset, \epsilon, \mathcal{B}'(x)\right\rangle\} \  \cap $  \texttt{Nogood}$= \emptyset$, and \\
   &  & \hspace{0.4cm} (b) $\left\langle \mathcal{C}(x), R, \mathcal{B}(x)\right\rangle \notin$ \texttt{Nogood}\\
   & then &   choose $\mathcal{B}(x)$ as the current propositional branch of $x$ \\

          $LII$-rule: &  if & 1. $x$ is not blocked, and \\
          &  & 2. there are (modal constraints on $R$) $\exists^{\triangleright\!\triangleleft n} R.C \in \mathcal{B}'(x)$, and  \\
          &  & 3. $x$ has no LII for those modal constraints on $R$ \\
   & then &   generate an LII for those modal constraints on $R$ in $\mathcal{B}'(x)$, and \\
   &  &   generate upto $2^\lambda-1$ atom-decompositions as $R$-successors \\ \hline

    \end{tabular}

  \vspace{0.1cm}

\hspace{0.5cm}  Fig-1. The tableaux expansion rules\footnote{For clarity, we purposely do not show GCIs in these rules. However, the rules and the algorithm must take the chunk GCI into consideration.} for $\mathcal{ALCQI}$

  \vspace{0.1cm} 
  
The atom-decomposition for a set of \emph{modal constraints} on a certain role generates all possible combinations about \emph{role fillers} or \emph{negated role fillers}. Each combination is considered as conjuncted together. Also see footnote \ref{x}. For example, for the set $\{\exists^{\leq 3}R.C_1, \exists^{\geq 2}R.C_2, \exists^{\geq 4}R.C_3\}$ of \emph{modal constraints} on role $R$, the \emph{atomic decomposition} is $\{C_1 \sqcap C_2 \sqcap C_3; C_1 \sqcap C_2 \sqcap \neg C_3; C_1 \sqcap \neg C_2 \sqcap C_3; C_1 \sqcap \neg C_2 \sqcap \neg C_3; \neg C_1 \sqcap C_2 \sqcap C_3; \neg C_1 \sqcap C_2 \sqcap \neg C_3; \neg C_1 \sqcap \neg C_2 \sqcap C_3\}$ of $2^3-1$ elements. 

Given a completion structure, a node $x$ is blocked if none of its ancestors are blocked, and it has a \emph{witness} $x'$ such that

\begin{itemize}
\item $\mathcal{B}(x) = \mathcal{B}(x')$ and $\mathcal{B}'(x) = \mathcal{B}'(x')$
\end{itemize}

In this case, we say $x'$ blocks $x$. It is static and is based on propositional-branch equality. For details see below on soundness and completeness.

\vspace{0.1cm}
The \emph{primitive clashes} $\mathcal{ALCQI}$ include any superset of $\{\neg \top\}$, $\{C, \neg C\}$, and $\{\exists^{\leq -1} R.C\}$. The latter is new and is for \emph{fine-tuned modal constraints}. It is reasonable to require that the constants in modal constraints (i.e. qualified number restrictions) are given as non-negative integers. By fine-tuning, possibly it gets a constraint like $\{\exists^{\leq -1} R.C\}$ which we stipulate as trivially unsatisfiable.

To generalize primitive clashes, we use the $\bot$-sets originally introduced in \cite{ExpALC2000}. Inconsistency inference is performed on demand by tableau procedures.

\vspace{0.1cm}
The following are the \emph{inconsistency propagation rules} for $\bot$-set.

\vspace{0.1cm}

        \begin{tabular}{clll} \hline

    $\bot$-0-rule: & &    &   $\{\neg \top\} \in \bot$-sets.\\

    $\bot$-1-rule: & &   &   $\{C, \neg C\} \in \bot$-sets.\\

    $\bot$-2-rule: & &    &   $\{\exists^{\leq -1}R.C\} \in \bot$-sets.\\

  $\bot$-3-rule: & &  if &   $\alpha \cup \{G, \mathcal{K}_a\} \in \bot$-sets \\
  & &   then &   $\alpha \in \bot$-sets. \\

  $\bot$-4-rule:& &   if &  (1) $\alpha\in \bot$-sets, and \\
  & & &  (2) $\alpha \subseteq \beta$ \\
  & &   then & $\beta \in \bot$-sets.\\\

  $\bot$-5-rule: & &   if &  (1) $\alpha \cup \{C\}\in \bot$-sets, and \\
  & & &  (2) $\alpha \cup \{D\} \in \bot$-sets  \\
  & &  then &  $\alpha \cup \{C \sqcup D\}\in \bot$-sets.\\ 

  $\bot$-6-rule: & & if &  (1) the set of modal constraints about $R$ is $\mathcal{M}$, and \\
    & & &  (2) $\mathcal{M}$'s atom decompositions about $R$-role-fillers is $\mathcal{D}$, and \\
    & & &  (3)  $\mathcal{D}$'s linear-integer-inequalities $lii$ is infeasible\\
    & &   then &  $\mathcal{M} \in \bot$-sets.\\ \hline

    \end{tabular}

  \vspace{0.1cm}

  \hspace{0.5cm}    Fig-2. The inconsistency propagation rule for $\mathcal{ALCQI}$

  \vspace{0.2cm}

For jargons in LP/IP, see \cite{linbook} and \cite{schrijver86}. 

Here is the outline\footnote{It will not be presented in this paper due to space limit. For details see \cite{Ding2007b}.} of the intended decision procedure. The decision procedure uses a \emph{restart strategy}\footnote{The use of restart here is for an easy presentation of the complexity argument.}  and takes a depth-first traversal to construct a tableaux tree. It uses two global data structures. \texttt{Nogood} permanently  holds triplets like $\left\langle \mathcal{C}(x), edge2me, \mathcal{B}(x)\right\rangle$, $\left\langle \mathcal{C}(x), edge2me, \mathcal{L}(x)\right\rangle$, and $\left\langle \emptyset, \epsilon, \mathcal{B}(x)\right\rangle$ for $\bot$-sets encountered. \texttt{Witness} holds intermediate results like $\left\langle \mathcal{B}(x), \mathcal{B}'(x)\right\rangle$, and is used for  \emph{blocking}. The restart strategy resets \texttt{Witness} to empty whenever $\bot$-rules can infer a new \texttt{Nogood} element bottom-up. This inconsistency inference is triggered by the primitive clashing or by the (cache) hitting of \texttt{Nogood}.

The procedure decides $E$ as \textsl{unsatisfiable} if $E \in  $ \texttt{Nogood};  or otherwise decides $E$ \textsl{satisfiable} if the size of \texttt{Nogood} is not changed. In other cases, the procedure restarts over and over. The termination is guaranteed since the size of \texttt{Nogood} is bounded and each restart will find a new (nontrivial) inconsistency set.

\section{Correctness}

\subsection{Completeness}

For the completeness, we need to prove the correctness for what regards concept unsatisfiability. Taking the approach in \cite{ExpALC2000}, we start with a lemma saying that $\bot$-rules correctly propagate inconsistencies. 

 \begin{lemma} The $\bot$-rules generate only unsatisfiable sets. \end{lemma}

 \begin{proof} By induction on the application of $\bot$-rules.

 \vspace{0.1cm}

 Base cases. Consider rules $\bot$-0, $\bot$-1, and  $\bot$-2. They are clearly unsatisfiable.

 \vspace{0.1cm}

 Inductive cases. Suppose the claim holds for the antecedent of each $\bot$-rule. We analyze the application of each $\bot$-rule.

\begin{itemize}
\item  ($\bot$-3):  Give $C$ is unsatisfiable w.r.t. $G$ and $\mathcal{K}_a$. Consider that $\top \sqsubseteq G$ and  $\top \sqsubseteq \mathcal{K}_a$, in every model for both $G$ and $\mathcal{K}_a$, $G$ and $\mathcal{K}_a$ are equivalent to $\top$. Then it is clear that $C$ is unsatisfiable.
\item  ($\bot$-4):  We prove the claim by contradiction. Suppose $\alpha \subseteq \beta$, $\alpha$ is unsatisfiable and $\beta$ is satisfiable. Let $M$ be a model for $\beta$. Using the sub-model generating technique, there is a sub-model $N$ of $M$ satisfies $\alpha$, and this contradicts the hypothesis that $\alpha$ is unsatisfiable.
\item  ($\bot$-5):  We prove the claim by contradiction. Suppose $\alpha \sqcap C$ and $\alpha \sqcap D$ are unsatisfiable, but $\alpha \sqcap (C \sqcup D)$ is satisfiable. Let $M$  be a model for $\alpha \sqcap (C \sqcup D)$, then either $\alpha \sqcap C$ or $\alpha \sqcap D$ is satisfied in $M$. This contradicts the hypothesis.
\item  ($\bot$-6):  The \emph{atom-decomposition} exhaustively generates all combinations of (negated) role fillers on one role $R$. The \emph{column vector} of the \emph{coefficient matrix} of $lii$ takes a value \textbf{0} if its corresponding role-filler combination is found \textsl{unsatisfiable}; otherwise it remains its initial value. We prove the claim by contradiction. Suppose $\mathcal{M}$ is satisfiable, then this leads to a feasible (conjuncted) combination of role fillers. This contradicts the hypothesis.  $\qed$ \end{itemize} \end{proof}

 \begin{lemma} \textbf{\emph{(Completeness)}}  If $n \in \bot$-sets, then $n$ is unsatisfiable.  \end{lemma}
 
\subsection{Soundness}

 Denote \textbf{T} the completed tree constructed. For node $x_i \in $ \textbf{T}, denote its initial label as $\mathcal{L}(x_i)$, its current propositional branch as $\mathcal{B}(x_i)$, and the fine tuned one as $\mathcal{B}'(x_i)$. The algorithm takes a DFS traversal to build \textbf{T} starting from the root node $x_0$, and uses the global data structures \texttt{Witness} and \texttt{Nogood}.

 We denote $x_i \triangleleft  x_j$ if $x_i$ is expanded (completed) before $x_j$ does.  The blocking relationship conforms to this (node expansion) ordering.   Only completed propositional branches enter their pairwise label sets in
 \texttt{Witness}. The blocking nodes must be \emph{propositionally completed} (so that the conventional $\sqcap$-rule and $\sqcup$-rule are no longer
 applicable.), fine-tuned and not in \texttt{Nogood}. 

 \begin{lemma} \textbf{\emph{(Soundness)}} If there is tableau tree \textbf{T} for $\mathcal{L}(x_0) = \{E\}$ w.r.t.  $G$ and $\mathcal{K}_a$, then there is a model $M$ for $\mathcal{L}(x_0)$ w.r.t.  $G$. \end{lemma}

 \begin{proof} It takes three steps.

(1) To admit infinite models, we consider paths in \textbf{T}. The mapping \textbf{Tail}($p$) returns the last element in a path $p$. Give a path $p = [x_0, ..., x_n]$, where $x_i$ are nodes in \textbf{T},  \textbf{Tail}($p$)$ = x_n$.   Paths  in \textbf{T} are defined inductively as follows:
\begin{itemize}
\item for  the root node $x_0$ in \textbf{T}, $[x_0]$ is a path in \textbf{T}.
\item for a path $p$ and a node $x_i$ in \textbf{T}, $[p,x_i]$ is a path in \textbf{T} iff
\begin{itemize}
\item $x_i$ is not blocked, and
\begin{itemize}
\item $x_i$ is a successor of \textbf{Tail}$(p)$ and the unknown\footnote{Each tableaux node corresponds to one variable of one $lii$ at its predecessor node.} $v_{x_i} > 0$, or
\item $y$ is a successor of \textbf{Tail}$(p)$ and $x_i$ blocks $y$ and the unknown $v_y > 0$.
\end{itemize}
\item $x_i$ is not known to be unsat (i.e., its related triplets $\notin$ \texttt{Nogood}), and 
\end{itemize} 
\end{itemize}

 The pre-model $M' = (\Delta^\mathcal{I'}, .^\mathcal{I'})$ can be defined with:
  
 $\Delta  = \{x_p | \  p $ is a path in \textbf{T} $\}$
 
 $x_p \in (\mathcal{L}($\textbf{Tail}$(p)) \sqcap \mathcal{B}($\textbf{Tail}$(p)) \sqcap \mathcal{B}'($\textbf{Tail}$(p)) )^\mathcal{I}$
 
 $\{\left\langle x_p, x_q\right\rangle | \left\langle x_p, x_q\right\rangle \in (R)^\mathcal{I} \}= \{ \left\langle x_p, x_q\right\rangle \in \Delta  \times \Delta  | $  $q = [p, $ \textbf{Tail} $(q)]$ and 
  
 \hspace{1cm} 1. \textbf{Tail}$(q)$ is an $R$-successor of \textbf{Tail}$(p)$, or
 
 \hspace{1cm} 2. $\exists y \in $ \textbf{T}, $y$ is an $R$-successor of \textbf{Tail}$(p)$ and \textbf{Tail}$(q)$ blocks $y \ \}$ 
 
 \hspace{0.6cm} $ \bigcup \ \ \ \  \{ \left\langle x_p, x_q\right\rangle \in \Delta \times \Delta | $  $ p = [q, $ \textbf{Tail} $(p)]$ and
 
  \hspace{1cm} 1. \textbf{Tail}$(p)$ is an $R^-$-successor of \textbf{Tail}$(q)$, or
 
 \hspace{1cm}  2. $\exists y \in $ \textbf{T}, $y$ is an $R^-$-successor of \textbf{Tail}$(q)$ and \textbf{Tail}$(p)$  blocks $y \ \}$ 
 
 (2) Consider the unknown variable $v_x$ that corresponds to each node $x$ of $M'$,
 duplicate as many $v_x > 0$ numbers of $x$ as the solution requires. This lead to the
 model  $M'' = (\Delta^\mathcal{I''}, .^\mathcal{I''})$.  Each element of $M''$ is clash-free
 and is saturated w.r.t. the local cardinality restrictions. $M''$ is a model for $E$ and $G$ and $\mathcal{K}_a$.

(3)  Use the \emph{sub-model generating technique} to
extract a model $M$ (for $E$ and $G$) from $M''$ (which is for $E$ and $G \cup \mathcal{K}_a$). $\qed$\end{proof}

\section{Complexity}

\begin{lemma} \textbf{\emph{(Termination)}} The algorithm terminates in $c^{O(n)}$ for some constant $c > 1$, where $n$ is the size of the converted problem. \end{lemma}
\begin{proof} (1) Due to the blocking strategy, the tree size is bounded by $a^{O(n)}$ for some constant $a > 1$. (2) Each node of the tree takes a single exponential cost in $n$. (3) The size of \texttt{Nogood} is bounded by another single exponential function in $n$. The restart strategy forces at least one new \texttt{Nogood} will be inferred when restarting happens. This guarantees at most $\|$\texttt{Nogood}$\|$ trees will be constructed. The termination is within  $c^{O(n)}$ for some constant $c > 1$. $\qed$ \end{proof}

\begin{theorem} The  tableau-based decision procedure decides $\mathcal{ALCQI}$ concept satisfiability problems in ExpTime in the worst case w.r.t. GCIs. \end{theorem}

\section{Summary and Related Work}

We have investigated the satisfiability problem in
$\mathcal{ALCQI}$ w.r.t. a set of general inclusion axioms and also the
applicability problem of the \emph{tableaux caching} technique in \emph{tree
structures} restricted by local cardinality constraints and
inverse relations. The work is inspired by the ExpTime
tableaux procedure given in ~\cite{ExpALC2000}. The topic of
tableaux-based reasoning for  qualified number restrictions has
been well investigated, and it requires a thorough study to distill
the contributions as previously made in
\cite{Ohlbach99} \cite{Tobies99} \cite{Haarslev01comb}
\cite{HoSa02a}, and \cite{BaaHlaLutWolLPAR03} \cite{HladikIJCAR04}, and many more on reasoning of finite models.  \cite{HoST00b} shows  that the
$\mathcal{SHIQ}$ enjoys the \emph{tree model property}, and so does the $\mathcal{ALCQI}$.

 We have blurred the distinction between the \emph{blocking technique} and the \emph{tableaux caching}. Regardless of the differences, both are for the termination of tableau procedures. The soundness issue of \emph{tableaux caching} come to the surface with inverse roles for years. There was a tackling of this problem\cite{towa2005} with the \emph{precompilation technique}. For an ExpTime procedure on  $\mathcal{ALCFI}$, see \cite{Ding2007b}.

In summary, we have presented (1) the use of the (restricted) analytic-cut for $\mathcal{ALCQI}$, and
(2) a tableau-based method of worst-case 
ExpTime insensitive to the coding of numbers, and (3) a way to use the
tableaux caching technique for a logic having both inverse roles and
qualified number restrictions w.r.t. GCIs. For a verbose version giving details of the algorithm see \cite{Ding2007b}.
Refinements, empirical issues and optimisations are to be considered in our
next work.

\section*{Acknowledgements}
The author thanks professor Vasek Chvatal for several inspiring discussions.

\bibliography{reference}
\bibliographystyle{alpha}
\clearpage

\end{document}